\documentclass[osajnl,twocolumn,showpacs,10pt]{revtex4-1}
\usepackage{amsmath,amssymb,graphicx}

\begin{document}

\title{Fabrication of a centimeter-long cavity on a nanofiber for cavity QED 
}
\author{Jameesh Keloth, K. P. Nayak, and K. Hakuta}

\begin{abstract}
We report the fabrication of a 1.2 cm long cavity directly on a nanofiber
using femtosecond laser ablation. The cavity modes with finesse value in the
range 200-400 can still maintain the transmission between 40-60\%, which can
enable \textquotedblleft strong-coupling\textquotedblright\ regime of cavity
QED for a single atom trapped 200 nm away from the fiber surface. For such cavity
modes, we estimate the one-pass intra-cavity transmission to be 99.53\%.
Other cavity modes, which can enable high cooperativity in the range 3-10,
show transmission over 60-85\% and are suitable for fiber-based single
photon sources and quantum nonlinear optics in the \textquotedblleft Purcell\textquotedblright regime. 
\end{abstract}

\maketitle

%\email{Corresponding author: kali@cpi.uec.ac.jp}

\address{Center for Photonic Innovations, University of
Electro-Communications, Tokyo 182-8585, Japan}

%\ocis{(350.4238) Nanophotonics and photonic crystals; (060.5565) Quantum
%communications.} % REPLACE WITH CORRECT OCIS CODES FOR YOUR ARTICLE
% NOTE: \ocis{} IS ALIASED TO \pacs{} BUT MUST
% FORMAT THE TERMS CORRECTLY FOR EACH JOURNAL

%% required

Efficient quantum state transfer between single photons and single atom is a
key challenge towards realization of quantum networks \cite{kimble1}.
Interaction of a single atom with strongly confined photons in an optical
cavity leading to cavity QED effects, is a promising approach to realize a
quantum interface \cite{kimble1,berman,rempe review}. A crucial requirement
to achieve strong interaction between single photon and single atom, is that
the single atom cooperativity parameter $C$ $=(2g_{0})^{2}/(\kappa \gamma
_{0})>>1$, where $2g_{0}$ is the single photon Rabi frequency, $\kappa $\ is
the cavity decay rate (linewidth) and $\gamma _{0}$ is the atomic spontaneous emission
rate in vacuum. Even with $C>>1$, there are two regimes with different
dynamics, a) \textquotedblleft Purcell\textquotedblright\ regime, when $%
\kappa >2g_{0}, \gamma _{0}$ and b) \textquotedblleft strong-coupling\textquotedblright\
regime, when $2g_{0}>\kappa, \gamma _{0}$. The \textquotedblleft
strong-coupling\textquotedblright\ regime has been investigated using
free-space Fabry-Perot (FP) cavities, where the coherent quantum phenomena
like single-atom lasing and vacuum Rabi oscillations have been demonstrated 
\cite{kimble2,kimble3,rempe review}. However, it requires extremely high
finesse of 0.3 to 0.5 million, which is technically challenging. Although
high quality mirrors with transmission and scattering loss less than 2 ppm
has been reported, but the overall cavity transmission may drop to 10-20\% 
\cite{rempe review,rempe cavity transmission}. Following the development of
free-space FP cavities, various designs of nanophotonic cavities have also
been developed and investigated. In particular, the designs have focused on
the \textquotedblleft Purcell\textquotedblright\ regime, for applications
like single photon generation, single photon switching and quantum nonlinear
optics, where high transmission is essential \cite{rempe review, lukin}.

In the vision of a quantum network, efficient integration of the quantum
interface to the existing fiber network is also an essential requirement. In
this context, optical nanofiber based cavities offer a flexible alternative
platform \cite{famsan2}. The nanofiber is the subwavelength diameter waist
of a tapered single mode optical fiber. Using adiabatic tapering condition 
\cite{adiabatic}, efficient mode coupling to nanofiber region can be
realized \cite{orozco}, enabling efficient integration to fiber networks.
Strong transverse confinement of guided fields down to wavelength scale can
be realized in the nanofiber. Moreover a major part of the guided field
propagates outside the fiber enabling interaction with the surrounding
medium. In order to get insight about the interaction dynamics in a
nanofiber based cavity we follow the formalism developed in ref. \cite%
{famsan2}. Based on the formalism, $2g_{0}=2\sqrt{\eta \gamma c/L}$ and $%
\kappa =\pi c/(FL)$, respectively, where $\eta $ is the channeling
efficiency of spontaneous emission of atom into the nanofiber guided modes
without a cavity, $\gamma $ is the atomic spontaneous emission rate near the
nanofiber, $c$ is the speed of light in vacuum, $L$ is the optical length of
the cavity and $F$ is the finesse of the cavity mode. From this one can get, 
$C=(2g_{0})^{2}/(\kappa \gamma _{0})\simeq 4\eta F/\pi $. One should notice
that the $C$ is independent of $L$ and mainly depends on $F$ and the
transverse confinement of the optical mode through $\eta $. The effective
mode waist of the nanofiber or other nanophotonic structures can be less than
1 $\mu $m which is one order smaller compared to the typical mode waist of
10 - 30 $\mu $m for free space FP cavities. As a result high cooperativity
can be achieved even for moderate finesse of 50 - 100. Furthermore in case
of nanofiber cavities one can independently control the cavity length to
reach the \textquotedblleft strong-coupling\textquotedblright\ regime since
the $\kappa $ value reduces faster than the $2g_{0}$ value as the cavity
length increases.

Fabrication of cavity structures on the nanofiber using the focused ion beam
milling has been demonstrated \cite{FIB}. Also fabrication of the photonic
crystal nanofiber (PhCN) cavity using the femtosecond laser ablation has
already been demonstrated \cite{phcn1,phcn2}. A composite photonic crystal
nanofiber cavity is also demonstrated by mounting a nanofiber on a
nanofabricated grating structure and using single quantum dot on such a
cavity, Purcell enhancement factor of 7 has been demonstrated \cite%
{composite1,compositeprl}. In the above cases, the cavity is formed directly
on the nanofiber and designed for operation in \textquotedblleft
Purcell\textquotedblright\ regime with high transmission of up to 80\%. The
cavity lengths ranged from 30 $\mu $m to few mm, depending on the cavity
design and mainly limited by the nanofiber length of few mm. On the other
hand, extremely long nanofiber cavities with cavity length of 10-33 cm are
also realized by splicing two conventional single mode fiber Bragg gratings
to the tapered fiber \cite{arno,aoki}. In this type of cavities, the
presence of the tapered section within the cavity may induce intra-cavity
loss and limit the achievable finesse and on-resonance transmission. The
one-pass intra-cavity transmission reported in ref. \cite{arno} and \cite%
{aoki} are 98.3\% and 94\%, respectively. Hence for a finesse in the range
50 - 100 the total cavity transmission may reduce to 10\% in ref. \cite%
{arno} and below 5\% in ref. \cite{aoki}. Nevertheless, using such extremely
long nanofiber cavity, strong-coupling between single trapped Cs-atoms and
the cavity guided photons have been demonstrated \cite{aoki}.

\begin{figure}[tbph]
\begin{center}
\includegraphics[width=8 cm]{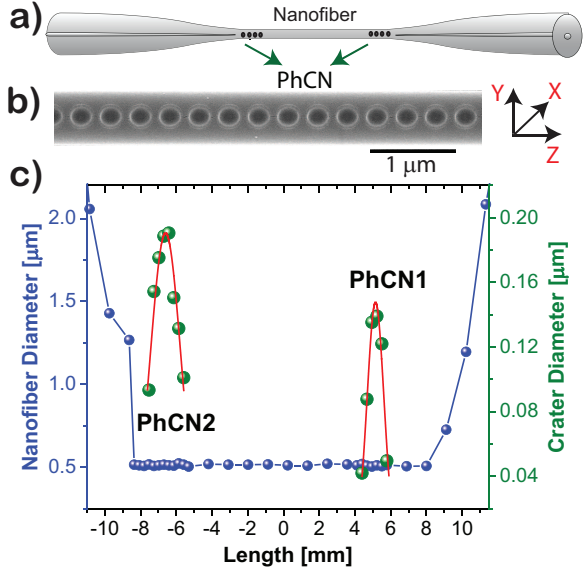}
\end{center}
\caption{(a) Schematic diagram of the long nanofiber cavity. (b) SEM image
of a typical fabricated PhCN sample. (c) Diameter profiles of the nanofiber
(blue circles) and the nano-craters (green circles) fabricated on it. The
red curves are the Gaussian fit to the nano-crater profiles. The blue line
is guide to the eyes. }
\label{fig1}
\end{figure}

In this letter, we report the fabrication of a centimeter-long cavity
directly on the nanofiber, which can operate both in \textquotedblleft
Purcell\textquotedblright\ and \textquotedblleft
strong-coupling\textquotedblright\ regimes of cavity QED. We demonstrate the
fabrication of a 1.7 cm long nanofiber with highly uniform diameter of 500 $%
\pm $ 2 nm over the entire length and maintaining high transmission of $>$%
99\%, which is a crucial requirement for this approach. Furthermore, we
fabricate two photonic crystal structures separated by 1.2 cm on such a
nanofiber using femtosecond laser ablation, thus forming a long nanofiber
cavity. The cavity modes with finesse value in the range 200-400 can still
maintain the transmission between 40-60\%, enabling \textquotedblleft
strong-coupling\textquotedblright\ regime for a single atom trapped 200 nm
away from the fiber surface \cite{famsan2}. For such cavity modes, we estimate the one-pass
intra-cavity transmission to be 99.53\%. Other cavity modes, which can
enable high cooperativity in the range 3-10, show transmission over 60-85\%
and are suitable for quantum nonlinear optics in the \textquotedblleft Purcell\textquotedblright regime. Moreover, placing solid-state quantum emitters directly on the nanofiber surface such cooperativity will be further enhanced by a factor of 5, which will be promising for fiber-based single photon sources. 

\begin{figure}[tbph]
\begin{center}
\includegraphics[width=8 cm]{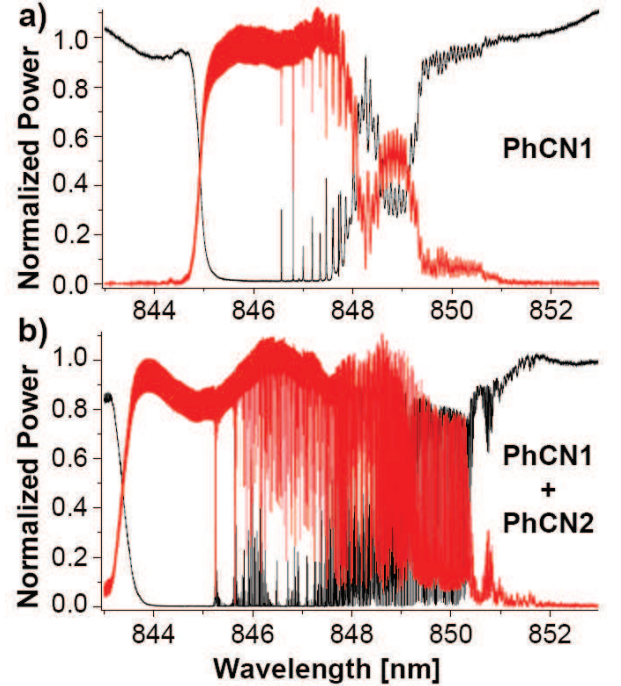}
\end{center}
\caption{Transmission (black lines) and reflection (red lines) spectra of
the nanofiber after the fabrication of the (a) first and (b) second PhCNs.
The spectra are measured for one of the principal polarizations (X-pol).}
\label{fig2}
\end{figure}

Figure 1(a) shows the schematic diagram of the nanofiber cavity. The
nanofiber is fabricated by tapering a standard single mode optical fiber
using heat-and-pull technique \cite{taper1}. The nanofiber is located at the
waist of the tapered optical fiber. In order to extend the nanofiber length
from few millimeters to few centimeters, we implement the linearly
increasing hot-zone technique \cite{taper1}. The profile of the tapered
fiber and the pulling parameters are designed based on the adiabatic
tapering guidelines detailed in ref. \cite{adiabatic,taper1}. By optimizing
the parameters we have realized $>$99$\%$ transmission for tapered
fibers with a total length of 7 to 8 cm, with a nanofiber waist diameter of 500 nm and waist length of 1.5 to
2.5 cm. The uniformity of the diameter over the entire length of the
nanofiber is $\pm $ 2 nm which is measured using a nondestructive and\textit{%
\ in situ} method detailed in ref. \cite{diameter}. The photonic crystal
structures are fabricated at the two ends of the nanofiber using the
femtosecond laser ablation \cite{phcn1,phcn2}, thus forming a long nanofiber
cavity. The scanning electron microscope image of a typical part of the PhCN
is shown in Fig. 1(b). One can see that periodic nano-crater structures with a period of 350 nm are
fabricated on the nanofiber \cite{phcn1, phcn2}. Figure 1(c) shows the
diameter profile of a long nanofiber cavity whose optical properties are
discussed in the following paragraphs. It shows the diameter profile of the
nanofiber (blue circles) and that of the nano-craters (green circles)
fabricated on it. The nanofiber diameter is 500 nm and it is uniform over a
length of 1.7 cm. The two PhC structures (indicated as PhCN1 and PhCN2) are
fabricated on the nanofiber with a separation of 1.2 cm. The PhCN1 is
fabricated first and the PhCN2 is fabricated later. Each PhCN structure
consists of thousands of periodic nano-craters \cite{phcn1, phcn2}. The
diameter profiles of the nano-craters show a peak-like structure. The
diameters at the peak of the profiles for the PhCN1 and PhCN2 are 140 nm and
190 nm, respectively. The red curves show the Gaussian fits to the diameter
profiles yielding the 1/e$^{2}$-widths of 0.9 mm and 1.7 mm for the PhCN1
and PhCN2, respectively. This suggests that the second fabrication was
stronger than the first one.

The transmission and reflection spectra of the long nanofiber cavity are
measured using a tunable, narrow linewidth diode laser (Newport TLB6700).
The laser is coupled to the tapered fiber through a 99:1 fiber-inline beam
splitter. The input light is launched through the 1\% port and the
transmission is measured after the tapered fiber, whereas the reflection is
measured through the 99\% port in the reverse direction. The spectra are
recorded by monitoring the power in the transmission and reflection ports
using photodiodes while the laser frequency is being scanned. The
polarization control is achieved using a fiber-inline polarizer before the
tapered fiber. Figure 2(a) shows the transmission (black curve) and
reflection (red curve) spectra after the fabrication of the PhCN1 for the polarization
perpendicular to the nano-crater faces (X-pol). One can
see that the stopband extends over 845 nm to 848 nm, where the light is
strongly reflected back. Also one can notice that sharp cavity modes appear
in the red-side of the stopband. The mode spacing is 95.5 GHz corresponding
to a cavity length of 1.3 mm.\ As discussed in the ref. \cite{phcn1, phcn2},
those cavity modes are due to the apodized index variation along the PhCN1.
Figure 2(b) shows the transmission (black curve) and reflection (red curve)
spectra after the fabrication of both the PhCN1 and PhCN2. One can see that
the width of the stopband increased and extends over 844 nm to 850 nm.
Moreover many closely spaced cavity modes appeared. This suggests that the
second fabrication was stronger and confirms proper overlap between the
stopbands of PhCN1 and PhCN2 to form a long nanofiber cavity.

\begin{figure}[tbph]
\begin{center}
\includegraphics[width=8 cm]{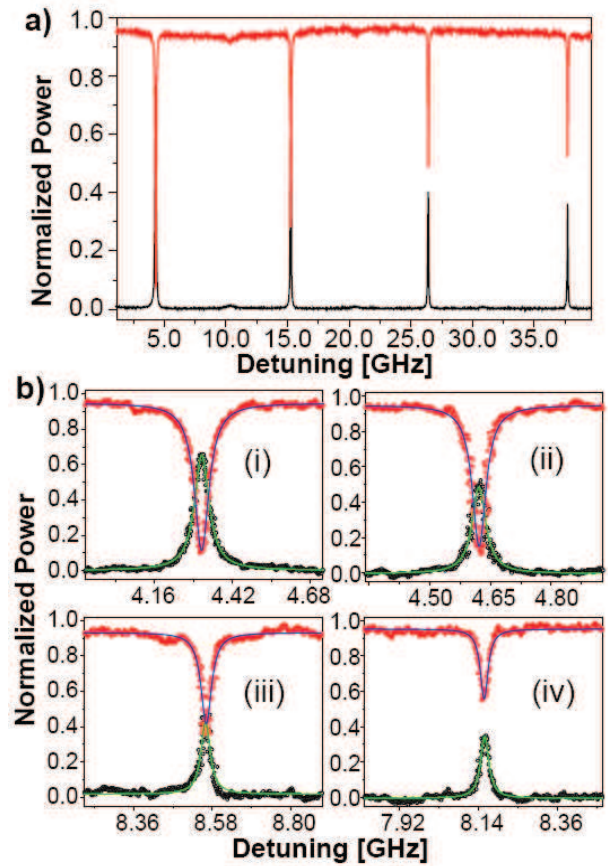}
\end{center}
\caption{(a) A typical part of the transmission (black curve) and reflection
(red curve) spectra of the long nanofiber cavity for X-pol, showing periodically spaced
cavity modes . (b) The transmission (black circles) and reflection (red
circles) spectra for four typical cavity modes (i-iv) shown individually.
The green and blue curves are the corresponding Lorentzian fits.}
\label{fig3}
\end{figure}

In order to properly resolve the cavity modes and precise normalization of
the on-resonance transmission and reflection values, we use a CW Ti-sapphire
laser source (MBR-110, Coherent Inc.). In this measurement the laser frequency is
locked to a reference cavity and the linewidth is 100 kHz. The spectra are
recorded by monitoring the power in the transmission and reflection ports
while stretching the tapered fiber. The details of such a technique was
reported in ref.\cite{phcn2}. Although we chose this method to be more
reliable, we must mention that the spectra measured using laser frequency
scanning technique yielded similar results with only marginal differences. A
typical part of the transmission (black curve) and reflection (red curve)
spectra for X-pol is shown in Fig. 3(a). One can see
periodically spaced sharp cavity modes. The mode spacing ($\Delta \nu _{FSR}$%
) is 10.36 GHz. From the $\Delta \nu _{FSR}$\ value we estimate a cavity
length ($l=L/n_{eff}=c/(2n_{eff}\Delta \nu _{FSR})$) of 1.2 cm, where $%
n_{eff}$ ($\simeq $1.2) is the effective index of the nanofiber guided mode. 
Four typical cavity modes are
shown individually in Fig. 3(b). The data for the measured transmission and
reflection spectra for the cavity modes are shown in black and red circles,
respectively. The green and blue curves show the Lorentzian fits. The
linewidths ($\kappa $) for the modes marked as i, ii, iii and iv are 59, 41,
33 and 27 MHz corresponding to the finesse ($F=\Delta \nu _{FSR}/\kappa $)
values of 175, 252, 314 and 384, respectively. One can notice that the
on-resonance transmission ($T_{0}$) and reflection ($R_{0}$) is increasing
and decreasing, respectively as the cavity linewidth is increased.

\begin{figure}[tbph]
\begin{center}
\includegraphics[width=8 cm]{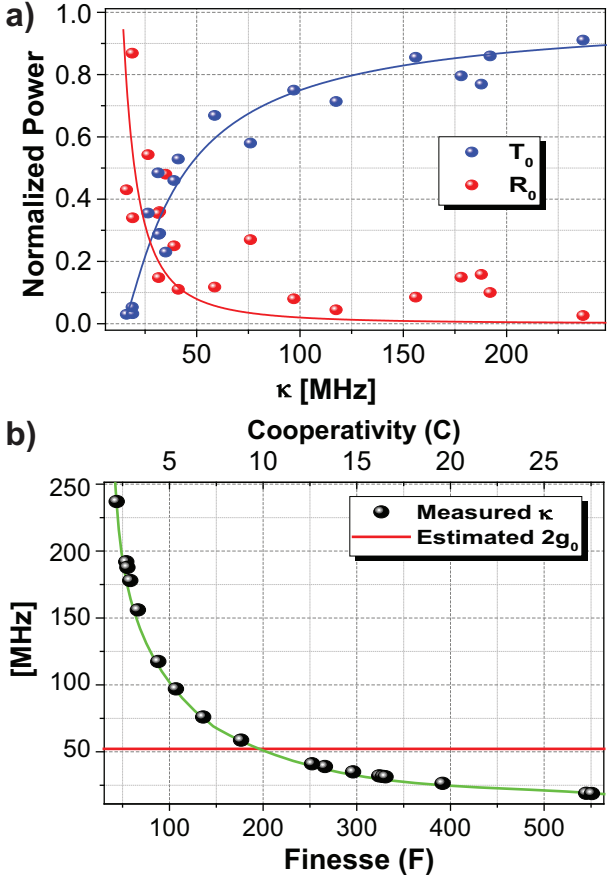}
\end{center}
\caption{(a) The measured on-resonance transmission, T$_{0}$ (blue circles)
and reflection, R$_{0}$ (red circles) values vs the cavity linewidths ($%
\protect\kappa $) for typical cavity modes. The blue and red lines are the
fits using Eq. 2. (b) The black circles show the measured cavity linewidths (%
$\protect\kappa $) vs the finesse ($F$) values for the typical cavity modes.
The green line is a fit showing the inverse relation. The red line shows the
estimated single photon Rabi frequency ($2g_{0}$), assuming a Cs-atom
trapped 200 nm away from the fiber surface. The corresponding single atom
cooperativity ($C$) is shown in the top axis.}
\label{fig4}
\end{figure}

The amplitudes of the field transmission ($t$) and reflection ($r$) of a
two-sided cavity can be formulated \cite{milburn} as

\begin{equation}
t=\frac{\sqrt{\kappa _{1}\kappa _{2}}}{\kappa /2+i\Delta \varpi };\text{ \ }%
r=\frac{\frac{1}{2}(\kappa _{1}-\kappa _{2}-\kappa _{s})-i\Delta \varpi }{%
\kappa /2+i\Delta \varpi }
\end{equation}

where $\Delta \varpi $ is the detuning between the laser frequency and the
cavity resonance, $\kappa _{s}$ is the intra-cavity loss rate, $\kappa _{1}$
and $\kappa _{2}$ are the coupling rates of the input and output side
mirrors, respectively and $\kappa =\kappa _{1}+\kappa _{2}+\kappa _{s}$ is
the cavity linewidth. The power transmission ($T$) and reflection ($R$) from
the cavity are given by $T=\left\vert t\right\vert ^{2}$ and $R=\left\vert
r\right\vert ^{2}$. Assuming a symmetric cavity ($\kappa _{1}=\kappa
_{2}=\kappa _{c}$) the on-resonance ($\Delta \varpi =0$) transmission and
reflection can be written as

\begin{equation}
T_{0}=\left\vert \frac{2\kappa _{c}}{\kappa }\right\vert ^{2}=\left\vert 1-%
\frac{\kappa _{s}}{\kappa }\right\vert ^{2};\text{ \ }R_{0}=\left\vert \frac{%
\kappa _{s}}{\kappa }\right\vert ^{2}.
\end{equation}%
From the above equation it is clear that the $T_{0}$ and $R_{0}$ values will
increase and decrease, respectively as the $\kappa $ increases. Moreover,
when the $T_{0}$ and $R_{0}$ values are equal, the total out coupling rate ($%
2\kappa _{c}$) and the intra-cavity loss rate ($\kappa _{s}$) balance each
other and one can get $\kappa _{s}=\kappa /2$.

Figure 4(a) shows the $T_{0}$ (blue circles) and $R_{0}$ (red circles)
values for the selected cavity modes (along with the modes shown in Fig.
3(b)) plotted against the corresponding $\kappa $\ values. In the selection
process we have chosen the cavity modes that have the highest transmission
for the corresponding $\kappa $\ values. The blue and red lines are the fits
using Eq. 2. From the fits we estimate the lowest $\kappa _{s}$\ value to be
15$\pm $1 MHz. Also one can clearly see from the plot that the $T_{0}$ and $%
R_{0}$ values are equal around 30 MHz. From this $\kappa _{s}$\ value, we estimate the one-pass
intra-cavity transmission to be 99.53\%. We have fabricated several cavity samples
and measured the lowest $\kappa _{s}$\ value to be in the range 15 - 20 MHz. This 
suggests that the fabrication process is reproducible. We must mention that we have 
also measured the cavity characteristics for the orthogonal polarization (Y-pol). We have
found that the cavity transmission for Y-pol is smaller compared to the X-pol, resulting in
higher $\kappa _{s}$\ value.

Based on the optical characteristics of the cavity we now estimate the
potential of the cavity in the context of cavity QED and quantum information
application. Figure 4(b) summaries the measured $\kappa $\ values
corresponding to the $F$ values. The estimated $C$ and $2g_{0}$ values are
also shown assuming a single Cs-atom trapped 200 nm away from the fiber surface \cite{famsan2}.
This shows that for the measured cavity modes we can achieve high
cooperativity ranging from 3 to 25. For cavity modes having $\kappa $\
values smaller than 50 MHz (i.e. finesse in the range 200 to 400), "strong-coupling" regime can be realized. As
shown in Fig. 4(a), the on-resonance transmission for such cavity modes can
be as high as 40$\%$ to 60$\%$. On the other hand, for the cavity modes
having $\kappa $\ values between 50 MHz to 170 MHz, the on-resonance
transmission range from 60$\%$ to 85$\%$, while maintaining a cooperativity
of 3-10. We must mention that using solid-state quantum emitters like
quantum dot or color centers in nanodiamonds, the emitter can be placed
directly on the nanofiber surface leading to much higher cooperativity in
the range 15-50 \cite{famsan2}. Hence such cavity modes can be implemented for fiber-based
single photon sources and quantum nonlinear optics in "Purcell" regime.

In conclusion, we have demonstrated the
fabrication of a 1.7 cm long nanofiber with highly uniform diameter of 500 $%
\pm $ 2 nm over the entire length and maintaining high transmission of $>$%
99\%. Furthermore, we
fabricate two photonic crystal structures separated by 1.2 cm on such a
nanofiber using femtosecond laser ablation, thus forming a centimeter-long
nanofiber cavity. The high optical quality of such a cavity shows promising
avenues for fiber-based quantum interface and single photon sources.

This work was supported by the Japan Science and Technology Agency (JST) as
one of the Strategic Innovation projects. KPN acknowledges support from a
grant-in-aid for scientific research (Grant no. 15H05462) from the Japan
Society for the Promotion of Science (JSPS).

\end{document}